\documentstyle[11pt,AATS,epsf]{article}	
\begin{document}	

\title{MACHO Project Analysis of the Galactic Bulge Microlensing Events with 
Clump Giants as Sources} 

\author{P.~Popowski$^1$, T.~Vandehei, K.~Griest, C.~Alcock, R.A.~Allsman, 
D.R.~Alves, T.S.~Axelrod, A.C.~Becker, D.P.~Bennett, K.H.~Cook, A.J.~Drake, 
K.C.~Freeman, M.~Geha, M.J.~Lehner, S.L.~Marshall, D.~Minniti, C.A.~Nelson, 
B.A.~Peterson, P.J.~Quinn, C.W.~Stubbs, W.~Sutherland,
D.~Welch (The~MACHO~Collaboration)}
\affil{$^1$ IGPP/LLNL; e-mail: popowski@igpp.ucllnl.org.}


\begin{abstract}
We present preliminary results of the analysis of 5 years of MACHO data
on the Galactic bulge microlensing events with clump giants as sources.
This class of events allows one to obtain robust conclusions because relatively
bright clump stars are not strongly affected by blending.
We discuss: 1) the selection of `giant' events, 2) the distribution of 
event durations, 
3) the anomalous character of event durations and optical depth in the MACHO 
field 104 centered on 
$(l,b) = (3 \hbox{$.\!\!^\circ$} 1, -3 \hbox{$.\!\!^\circ$} 0)$.
We report the preliminary average optical depth of 
\mbox{$\tau = (2.0 \pm 0.4) \times 10^{-6}$} (internal)
at $(l,b) = (3 \hbox{$.\!\!^\circ$} 9, -3 \hbox{$.\!\!^\circ$} 8)$, and
present a map of the spatial distribution of the optical depth.
When field 104 is removed from the sample, the optical depth drops to
$\tau = (1.4 \pm 0.3) \times 10^{-6}$, which is in excellent agreement
with infrared-based models of the central Galactic region.
\end{abstract}

\section{Introduction}

The structure and composition of our Galaxy is one of the 
outstanding problems in contemporary astrophysics. 
Microlensing is a powerful tool
to learn about massive objects in the Galaxy.
The amount of matter between the source and observer
is typically described in terms
of the microlensing optical depth, which is defined as the probability that a 
source flux will be gravitationally magnified by more than a factor of 1.34.
Early analyses (Udalski et al.\ 1994; Alcock et al.\ 1997) of the lines of 
sight toward the Galactic center produced two unexpected results: 1) a very
high optical depth of $3-4 \times 10^{-6}$ inconsistent with Galactic models 
and observations, 2) an overabundance of long events.
Here we analyze a new set of events to probe these controversial issues.

Blending is a major problem in any analysis of the 
microlensing data involving point spread function photometry.
The bulge fields are crowded, so that
the objects observed at a certain atmospheric seeing are blends of several 
stars, of which only one is typically lensed. This complicates
a determination of an event's parameters and the analysis of the detection 
efficiency of microlensing events.
If the sources are bright one can avoid these problems.
Red clump giants are among the brightest and most
numerous stars in the bulge.
Therefore, this analysis concentrates
on the events where the lensed stars are clump giants.

\section{The Data and Selection of ``Giant'' Events}

The MACHO Project observations were performed with $1.27$-meter telescope at 
Mount Stromlo Observatory, Australia.
In total, we collected 7 seasons (1993-1999) of data in the 94 Galactic 
bulge fields. The data that are currently available for the analysis 
consist of 5 seasons (1993-1997) in 77 fields, and
contain the photometry of about 30 million 
stars, including 2.1 million clump giants.

The events with clump giants as sources have been selected from
the sample of all events which contains about $\sim 280$ candidates.
The determination of which of these sources are clump giants, is investigated
through the analysis of the global properties of the color-magnitude
diagram in the Galactic bulge.

Using the accurately measured extinction towards Baade's Window 
allows one to locate {\it bulge} clump giants on the 
reddening-free color-magnitude diagram. 
This defines the parallelogram-shaped box in the upper left corner
of the left panel of Figure 1.
With the assumption that the clump populations in the whole bulge
have the same properties as the ones in the Baade's Window,
the parallelogram described above
can be shifted by the reddening vector
to mark the expected locations of clump giants in different fields.
The solid lines are the boundaries of the region where one could find
the clump giants in fields with different extinctions.
Using this approach, we identified 52 clump events.

\section{Field 104}

There is a high concentration of long-duration events in MACHO field 104 at
$(l,b) = (3 \hbox{$.\!\!^\circ$} 1, -3 \hbox{$.\!\!^\circ$} 0)$
as 5 out of 10 clump events longer than 50 days are in 104. 
The analysis of event durations {\em uncorrected} for 
efficiencies provides a lower limit on the difference between field 
104 and all the other fields.
We use the Wilcoxon's test 
on two samples: events in field 104 and all
the remaining ones, and find that
the events in 104 differ (are longer) at the level of $2.55 \sigma$.
In addition, field 104 also has the highest optical depth (see Figure 1).
Both of these features can be explained by the concentration of mass 
along this particular line of sight.

\section{Optical Depth}

First, we report the average optical depth of:
$$
\tau = (2.0 \pm 0.4) \times 10^{-6}\;\;\;\;\; {\rm at} \;\;\;\;\; (l,b) = (3 \fdg 9, -3 \fdg 8),
$$
which is 1.5-2 times lower than the previously obtained values.
We caution that this result is only preliminary, 
because possible systematic errors may be a fair fraction of the statistical
error.
We also note that about 40\% of the optical depth is in the events longer than 
50 days. This is at odds with standard models of the Galactic 
structure and kinematics.

Second, we plot in the right panel of Figure 1 the spatial distribution of 
the optical depth. 
The variation of the optical depth is dominated
by the Poisson noise. The gradient of the optical depth is stronger in $b$
than in $l$ direction.
Again, we note the anomalous character of field 104. It is marked
with a black square and has an optical depth of $(1.4 \pm 0.5) \times 10^{-5}$.

\begin{figure}[htb]	
\plottwo{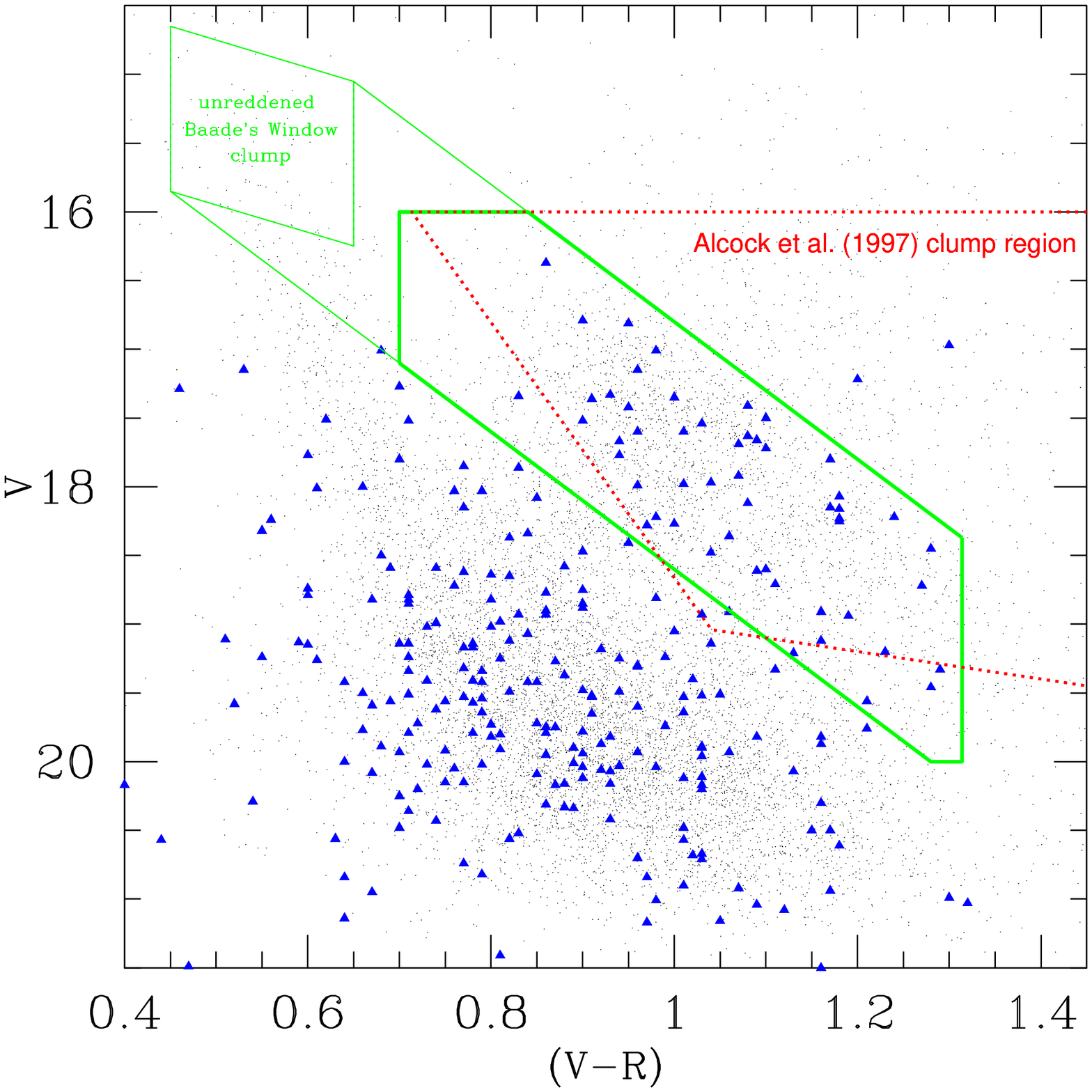}{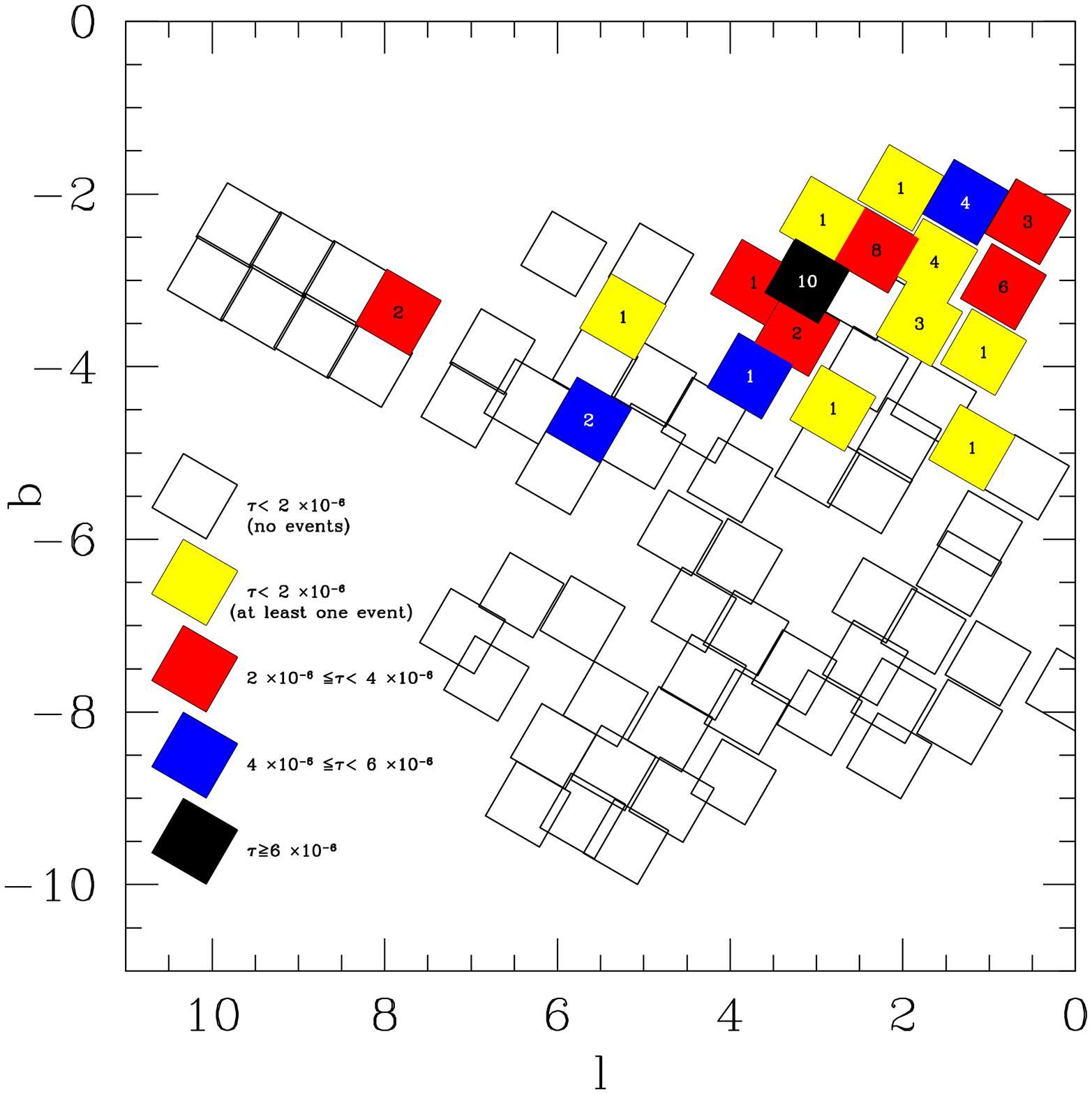}
\caption{{\em Left panel}:
Events are marked with filled triangles.
The region surrounded by a bold line is our clump region. 
For comparison, 
we plot with a dotted line the clump region from Alcock et al.\ (1997). 
Both selections return very similar events.
{\em Right panel}:
The spatial distribution of optical depth with the numbers of events given
in the center of each field.  Note anomalous field 104 with 10 events and 
optical depth of $(1.4 \pm 0.5) \times 10^{-5}$.
}
\end{figure}

\section{Conclusions}

We presented first results from the analysis of five years of microlensing
data toward the Galactic bulge collected by the MACHO Collaboration. 

It is possible to select an unbiased sample of clump events 
based only on an event's position on the color-magnitude diagram.
Ten out of 52 clump events have durations $>50$ days, which implies that 
$\sim 40$\% of the optical depth is in the long events.
This is surprising, because long events are the most likely a result
of the disk-disk lensing (Kiraga \& Paczy\'{n}ski 1994), 
while clump giants trace the bar rather than the inner disk 
(Stanek et al.\ 1994).
Field 104 centered on $(l, b) = (3 \fdg 1, -3 \fdg 0)$ is anomalous:
it has longer events and substantially higher optical depth.
Both effects can be explained simultaneously by a concentration of mass 
along this line of sight.
The optical depth averaged over the clump giants in 77 fields is
$\tau = (2.0 \pm 0.4) \times 10^{-6}$ at $(3 \fdg 9, -3 \fdg 8)$.
When anomalous field 104 is removed, the optical depth drops to
$\tau = (1.4 \pm 0.3) \times 10^{-6}$, which is fully consistent with
infrared-based models of the Galactic bar.

\end{document}